\documentclass[twocolumn,aps,prb,showpacs,preprintnumbers,amsmath,amssymb,superscriptaddress,longbibliography]{revtex4-2}

\usepackage{hyperref}
\usepackage{graphicx}
\usepackage{amsfonts}
\usepackage{diagbox}

\begin{document}
\title{Entangelment Entropy on Generalized Brillouin Zone}
\author{Zhenghao Yang}
\author{Chaoze Lu}
\author{Xiancong Lu}
 \email{xlu@xmu.edu.cn}
\affiliation{Department of Physics, Xiamen University, Xiamen 361005, China}

\begin{abstract}
  We investigate the entanglement properties of non-Hermitian
  Su-Schrieffer-Heeger (SSH) model from the perspective of the
  Generalized Brillouin Zone (GBZ). The non-Bloch entanglement entropy
  is defined on a quasi-reciprocal lattice, obtained by performing an
  ordinary Fourier transformation on the non-Bloch Hamiltonian. We
  demonstrate that the broken bulk-boundary correspondence is
  recovered in terms of the non-Bloch entanglement entropy. When the
  GBZ is circular, we show that the non-Bloch entanglement entropy is
  well-defined (real and positive-definite) in large parameter
  regions, except close to the exceptional points (EPs). In the
  critical region, we found that each Fermi point contributes
  precisely 1 to the central charge $c$ of the logarithmic scaling. At
  the EP, the central charge becomes negative due to the presence of
  the exceptional bound state. For the case of non-circular GBZ,
  long-range hopping emerges in the quasi-reciprocal lattice, and the
  von Neumann entropy on the GBZ is no longer real. However, the
  non-Bloch edge entanglement entropy remains real, which serves as a
  reliable topological indicator and respects the bulk-boundary
  correspondence.  We compute the topological phase diagram and reveal
  the critical behavior along the exceptional phase
  boundaries.
\end{abstract}
\maketitle

\section{introduction}

Non-Hermitian quantum mechanics naturally arises from the study of
open quantum systems coupling with an external
environment \cite{as.go.20,be.bu.21}.
For non-Hermitian systems with nonreciprocal hoppings, the bulk
eigenstates under open boundary conditions (OBC) accumulate at
boundaries, which is named by the non-Hermitian skin effect (NHSE)
\cite{ya.wa.18,ya.so.18a,ku.ed.18,le.th.19} and has been extensively
studied recently \cite{gh.da.19,zh.zh.22,li.ta.23,ba.sa.23}. In the presence of NHSE,
the energy spectrum sensitively depends on the boundary conditions:
the energy spectrum under OBC forms a star-shaped branching figures
\cite{ta.le.23}, while the energy spectrum under periodic boundary
conditions (PBC) makes up loops enclosing the OBC spectrum
\cite{zh.ya.20,ok.ka.20,da.kr.20}. The NHSE also results in the breakdown of
the celebrated bulk-boundary correspondence (BBC) of topological band
theory \cite{qi.zh.11,ha.ka.10}: the transition points of bulk
topological invariants do not match with transition points of zero
modes under OBC \cite{lee.16,xion.18,ya.wa.18}. In order to understand
the topology of non-Hermitian systems, the non-Bloch band theory and
the generalized Brillouin zone (GBZ) are proposed, based on which the
BBC for non-Hermitian systems can be reestablished
\cite{ya.wa.18,yo.mu.19,ya.zh.20}. Motivated by this breakthrough, the
non-Bloch band theory has been widely used, e.g., to construct Green
function \cite{xu.li.21}, to investigate the dynamical phenomena
\cite{long.19} and the non-Bloch parity-time symmetry \cite{xi.de.21}. 

Quantum entanglement offers a unique perspective on the exotic phases
and phase transitions of Hermitian quantum many-body systems
\cite{am.fa.08,ei.cr.10}. The degree of entanglement is quantified by
the entanglement entropy (EE), which can be obtained by partitioning
the system into several subsystems. Importantly, the topological EE is
the quantity to characterize the topological order of a gapped system
\cite{le.we.06,ki.pr.06}. For the one-dimensional (1D) Hermitian
critical (gapless) system, the scaling behavior of EE with the size
of subsystem contains valuable information about the critical
point, such as the central charge of the unitary conformal
field theory (CFT) \cite{ho.la.94,ca.ca.09}.

The entanglement properties of the non-Hermitian quantum systems is
also an interesting topic which has attracted much attentions recently
\cite{co.ja.17,he.re.19,ch.yo.20,le.le.20,mu.le.20,ch.ch.21,ok.sa.21a,ba.do.21,mo.ma.21,gu.yu.21,lee.22,tu.tz.22,ch.zh.22,fo.ar.23,ka.nu.23,xu.li.23,yi.ha.23,pa.wa.23,wa.je.23,or.he.23,hs.ch.23,or.im.23,ga.tu.23,li.yu.24,zhou.24,ji.do.24,li.li.24,wa.fa.24}. The
conception of EE has been generalized to the non-Hermitian systems in
the biorthogonal basis \cite{ch.yo.20,he.re.19}. A logarithmic scaling
with central charge $c=-2$ is found at critical point of the
non-Hermitian Su-Schrieffer-Heeger (SSH) model, which can be explained
by the $bc$-ghost non-unitary CFT \cite{ch.yo.20,tu.tz.22}. However,
the physical meaning of EE remains somewhat ambiguous in this
straightforward generalization \cite{tu.tz.22,fo.ar.23}: The
eigenvalues of reduced density matrix $\rho_A$ are no longer positive
definite, which leads to a negative or even complex EE. This is
inconsistent with the probability interpretation of a measurable
quantity \cite{tu.tz.22,fo.ar.23}. Some attempts have been made to
remedy this situation, for example, the generic entanglement entropy
$S_A=-\mathrm{Tr}(\rho_A\ln|\rho_A|)$ is proposed in
Ref. \cite{tu.tz.22}, the modified partial trace formalism is
introduced in Ref. \cite{co.ja.17}. However, more systematic analysis
is needed in this direction.

Inspired by the success of non-Hermitian topological band theory in
terms of GBZ, we may ask whether it is possible to construct an EE on
the GBZ that has proper behaviors? We address this question in this
paper for the non-Hermitian systems with skin effect. We begin with a
GBZ Hamiltonian and then make a Fourier transformation to
obtain a quasi-reciprocal lattice \cite{le.li.20}. The entanglement
entropy on GBZ is calculated by partitioning this quasi-reciprocal
lattice. The behaviors of various EE will be examined for two types of
GBZ (circular or non-circular). Our study may shed light on the
physical meanings of GBZ and non-Hermitian EE, and also open the
possibility to study the many-body physics based on GBZ.

This paper is organized as follows: In Sec.~\ref{section:bee},
we review the basic notions of biorthogonal entanglement entropy. In
Sec.~\ref{section:nhm}, we introduce the non-Hermitian SSH model and
the quasi-reciprocal lattice, on which the entanglement entropy is
calculated. In Sec.~\ref{section:cirGBZ}, we present the results of
entanglement entropy on the circular GBZ. In
Sec.~\ref{section:noncirGBZ}, we present the results of entanglement
entropy on the non-circular GBZ. Finally, a brief summary and discussion are provided in
Sec.~\ref{section:summary}.

\section{biorthogonal entanglement entropy}
\label{section:bee}

We consider a diagonalizable non-Hermitian Hamiltonian 
$\mathcal{H}=\sum_{i,j}c^\dagger_iH_{ij}c_j$ with $\mathcal{H}\neq
\mathcal{H}^\dagger$. Here,
$c_i^\dagger (c_i)$ is
the fermionic creation
(annihilation) operator (on lattice site $i$) satisfying the usual
anticommutation relation $\{ c_i,c_j^\dagger \} = \delta_{i,j}$, $\{
c_i,c_j\} = 0$, 
and $H_{ij}$ denotes the element of a hopping matrix. In the framework of
biorthogonal quantum mechanics \cite{brod.13}, the Hamiltonian
$\mathcal{H}$ has two types of eigenvectors
\begin{equation}
  \mathcal H |R_n \rangle = E_n |R_n\rangle, \quad
  \mathcal H^\dagger |L_n \rangle = E_n^* |L_n\rangle,
\end{equation}
where $|L_n \rangle$ and $ |R_n \rangle $ are referred to the left and
right eigenvectors, respectively, and are required to satisfy the
biorthonormality condition $\langle L_m| R_n \rangle = \delta_{mn}$.
The Hamiltonian $\mathcal{H}$ can be diagonalized as
$\mathcal{H}=\sum_{n}E_nd_{Rn}^\dagger d_{Ln}$, 
with the creation operator $d_{Rn}^\dagger$ 
($d_{Ln}^\dagger$) being related to the eigenstate $|R_n\rangle $
($|L_n\rangle $). Due to the non-Hermiticity,
$d_{Ln}^\dagger\ne (d_{Rn})^\dagger$, but the
following anticommutation relations still hold \cite{ch.yo.20,he.re.19,gu.yu.21}: 
$\{d_{Lm}^\dagger, d_{Rn} \} = \delta_{mn}$,
$\{d_{Lm}^\dagger, d_{Ln}^\dagger \}=\{d_{Rm}, d_{Rn}\}=0$.
The transformation between quasiparticle operators and the original
lattice operators is
\begin{eqnarray}
  d_{Rn}^\dagger &=& \sum_i c_i^\dagger V_{in}, \quad
                           V_{in}= \langle i | R_n \rangle \nonumber\\
  d_{Lm} &=& \sum_j W^\dagger_{mj} c_j, \quad
                           W_{jm} = \langle j | L_m \rangle
\end{eqnarray}
in which the transformation matrices $V$ and $W$ meet the condition
$W^\dagger V = 1$ (\textit{i.e.}, $W^\dagger=V^{-1}$).

When the energy of non-Hermitian systems is complex, there are
different ways to define the many-body ground state.
In this
paper, we construct the many-body ground states by filling up the
levels to Fermi energy according to the real part of the energy
\cite{le.le.20,ch.yo.20,he.re.19,gu.yu.21,ch.pe.22}:
\begin{equation}
  |\Omega_{L}\rangle=\prod_{\mathrm{Re}(E_m)<E_F}d_{Lm}^\dagger|0\rangle,
  \quad
  |\Omega_{R}\rangle=\prod_{\mathrm{Re}(E_n)<E_F}d_{Rn}^\dagger|0\rangle.
\end{equation}
The biorthogonal density matrix of the ground state is defined as
$\rho^{RL}=|\Omega_R\rangle\langle\Omega_L|$, which in general is
neither Hermitian nor positive-definite \cite{tu.tz.22}. If partitioning the total
system into two subsystems $A$ and $B$ and tracing out all degrees of
freedom in subsystem $B$, one can obtain the reduced density matrix
$\rho^{RL}_A= \mathrm{Tr}_B\rho^{RL}$. Following the definitions for Hermitian
system, the von-Neumann entanglement entropy is given by \cite{ch.yo.20,he.re.19,tu.tz.22}
\begin{equation}
  S_A = - \mathrm{Tr} \Big[ \rho^{RL}_A \log (\rho^{RL}_A) \Big],
\end{equation}
and the $\alpha$-order of R\'enyi entropy is
\begin{equation}
  S^\alpha_A=\frac{1}{1-\alpha}\log \mathrm{Tr}\Big[ (\rho^{RL}_A)^{\alpha}\Big]
\end{equation}
with $\lim_{\alpha\to1}S^\alpha_A=S_A$. The entanglement entropy can
be computed using the method of correlation matrix
\cite{pesc.03,ch.he.04,pe.ei.09}. For free non-Hermitian Fermions, 
the biorthogonal correlation matrix $C^A$ of subsystem $A$ is defined
as \cite{ch.yo.20,he.re.19,gu.yu.21}
\begin{equation}
  C^A_{ij} = \mathrm{Tr}(\rho_A^{RL}c_i^\dagger c_j)
  =\langle\Omega_{L}|c_i^\dagger c_j|\Omega_{R}\rangle
     =\sum_{m\in occ.} W^\dagger_{mi} V_{jm},
\end{equation}
in which the sites $i,j$ are restricted inside the subsystem $A$, and
the index $m$ labels the occupied energy levels.  The single-particle
entanglement spectrum is obtained from the eigenvalue spectrum
$\{\xi_l\}$ of the matrix $C^A$ under periodic boundary condition.
Based on this, the von-Neumann and R\'enyi entropy are given by
\begin{equation}
  S_A=-\sum_l(\xi_l\log(\xi_l)+(1-\xi_l)\log(1-\xi_l)),
\end{equation}
\begin{equation}
  S_A^\alpha=\frac{1}{1-\alpha}\sum_l\log((1-\xi_l)^\alpha+\xi_l^\alpha).
\end{equation}
The definition of edge entanglement entropy $S^\alpha_{\text{edge}}$ reads
\begin{equation}
  S^\alpha_{\text{edge}}=S^\alpha_{\text{OBC}}-\frac{1}{2}S^\alpha_{\text{PBC}},
\end{equation}
where $S^\alpha_{\text{OBC}}$ and $S^\alpha_{\text{PBC}}$ are
$\alpha$-th R\'enyi entropies calculated under OBC and PBC,
respectively.
Here, half of $S_{PBC}^\alpha$ is subtracted to eliminate the leading
terms of entanglement. Note that, in defining entanglement, there are
two cuts in the case of PBC but only one cut in the case of OBC
\cite{wa.xu.15,le.le.20}.

The non-Hermitian entanglement entropy is a useful tool for detecting
topological phase transitions and extracting information about edge
states \cite{wa.xu.15,le.le.20,ch.pe.22}.  In addition, it can reveal
critical behavior within a system \cite{ch.yo.20,gu.yu.21,lee.22}: by examining the scaling of
entanglement entropy with subsystem size, valuable information about
critical points and exceptional points can be obtained.
The
entanglement entropy can be fitted using the following universal formula
\cite{ca.ca.04,lafl.16},
\begin{eqnarray}\label{scaling}
S_A^\alpha(L_A) = \frac{c}{6} \Big(1+\frac{1}{\alpha}\Big)
    \ln \Big( \frac{L}{\pi} \sin \Big[ \frac{\pi L_A}{L}\Big] \Big)
    +a_\alpha,
\end{eqnarray}
in which $c$ is the central charge described by conformal field theory
(CFT), $a_\alpha$ is a non-universal constant, $L(L_A)$ is the
length of system (subsystem $A$), and $\alpha$ is the order of Renyi
entropy (for the von-Neumann entropy, $\alpha=1$). In Eq.
(\ref{scaling}), the total system length $L$ is fixed and subsystem
length $L_A$ is variable. Another way to fit the entanglement
entropy is to fix the ratio $L_A/L=1/2$ (equal bipartition),
\begin{eqnarray}\label{scaling2}
  S_A^\alpha(L)=\frac{c}{6} \Big(1+\frac{1}{\alpha}\Big)
    \ln (L) + b_\alpha,
\end{eqnarray}
in which $b_\alpha$ is also a non-universal constant. Both fitting
methods are used in this paper.

\section{Quasi-reciprocal Hamiltonian based on generalized Brillouin zone}
\label{section:nhm}

\begin{figure}[t]
  \includegraphics[width=0.98\linewidth]{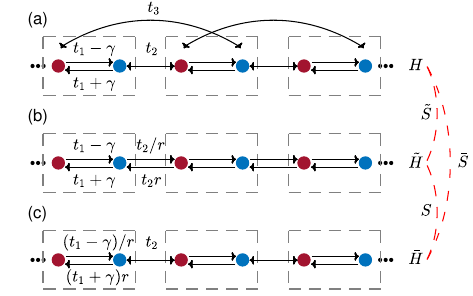}
  \caption{\label{fig:model}(a) Schematic diagram of the non-Hermitian
    SSH model. (b) The quasi-reciprocal model when $t_3=0$, obtained
    by applying the similarity transformation $\tilde{S}$ to the
    original non-Hermitian SSH model.  (c) The model after the
    similarity transformation $\bar{S}$. }
\end{figure}

\begin{figure}[t]
\includegraphics[width=0.98\linewidth]{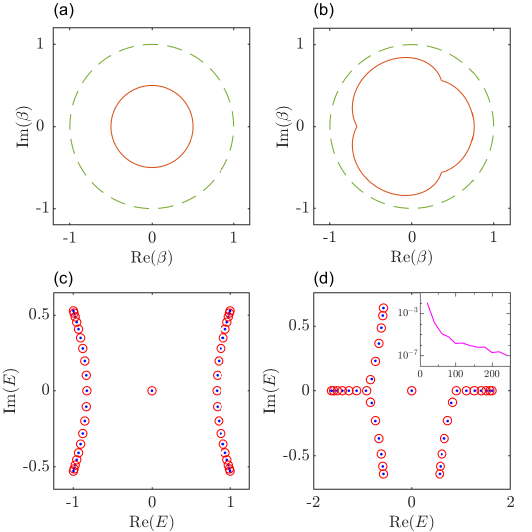}
\caption{The examples of circular GBZ (a) and non-circular GBZ (b).
  The green dashed circle denotes the conventional BZ with $|\beta|=1$.
  The radius of circular GBZ (orange circle) in (a) is everywhere
  smaller than $1$, indicating that wave functions only accumulate at
  the left end of the chain. Sub-figures (c) and (d) show the energy
  spectra of original OBC Hamiltonian $H_{OBC}$ (blue dots) and
  quasi-reciprocal OBC Hamiltonian $\widetilde{H}_{OBC}$ (red
  circles). The length of lattice is set to be $L=40$. For the case
  of circular GBZ (c), the blue dots precisely match the red circles.
  However, for the non-circular GBZ (d), the blue dots slightly
  deviate from the red circles, especially close to the
  bifurcations.
  The inset of sub-figure (d) shows the mean square error (vertical axis)
  between energy spectra of $H_{OBC}$ and $\widetilde{H}_{OBC}$ as a
  function of system length $L$ (horizontal axis). 
  We fix $t_3=0$ in (a) and (c), but $t_3=0.2$ in (b)
  and (d). Other parameters used in the calculations are $t_1=0.4$,
  $t_2=1$, and $\gamma=2/3$.}
\label{fig:gbz}
\end{figure}

We consider a typical non-Hermitian SSH model with the following Bloch
Hamiltonian \cite{ya.wa.18,lieu.18}
\begin{multline}\label{Hami}
  h(k)=\\\left(\begin{matrix}0& (t_1+\gamma)+t_2 e^{-\mathrm{i}k}+t_3
                                e^{\mathrm{i}k}\\(t_1-\gamma)+t_2
                 e^{\mathrm{i}k}+t_3
                 e^{-\mathrm{i}k}&0\end{matrix}\right),
\end{multline}
in which $k\in [0, 2\pi)$ is the crystal momentum. The hopping
processes of this model in real space are illustrated in
Fig. \ref{fig:model}(a). According to the theory of GBZ
\cite{ya.wa.18,yo.mu.19,ya.zh.20}, we substitute $e^{ik}$ with
$\beta=e^{ik'}$ ($k'=k+i\tau(k)\in \mathbb{C}$) to gain the
corresponding non-Bloch Hamiltonian
\begin{multline}\label{hbeta}
  h(\beta)=\\\left(\begin{matrix}0&
             (t_1+\gamma)+t_2\beta^{-1}+t_3\beta\\(t_1-\gamma)+t_2\beta+t_3\beta^{-1}
                                  &0\end{matrix}\right).
\end{multline}
The characteristic equation of $h(\beta)$ is given by
\begin{equation}
  \det[E-h(\beta)]=0,
\end{equation}
which is a quartic equation and has four solutions $\beta_i
(i=1,\cdots,4)$ for a specific energy $E$. If ordering the solutions by
their norms, $|\beta_1(E)|
\leqslant|\beta_2(E)|\leqslant|\beta_{3}(E)|\leqslant|\beta_{4}(E)|$,
the GBZ $C_\beta$ is determined by the condition \cite{ya.wa.18,yo.mu.19,ya.zh.20},
\begin{equation}\label{con_GBZ}
  |\beta_2(E)|=|\beta_3(E)|.
\end{equation}
The condition (\ref{con_GBZ}) implies that the two exponential
eigenstates, corresponding to $\beta_2$ and $\beta_3$, are of the same
order. This allows them to cancel each other near the boundaries of an
open lattice, forming a standing wave that satisfies the OBC
\cite{ya.wa.18,yo.mu.19}.
Two typical GBZs are shown in Fig. \ref{fig:gbz}(a) and (b), where the
GBZ forms a closed loop encircling the origin on the complex plane.
The OBC spectrum in the thermodynamic limit can be exactly
obtained when $\beta$ goes along the loop of GBZ.


To define the entanglement entropy in terms of GBZ, one needs to
construct an artificial real-space lattice which respects the
translation invariance.
Suppose that the phase angle of $\beta_{GBZ}$ is equally distributed on the
GBZ, we then construct non-Bloch Hamiltonian matrix $H(\beta_{GBZ})$ by
selecting the points whose phase angle belongs to the Brillouin zone (BZ)
of original physical lattices, that is,
\begin{equation}\label{km}
  \textrm{Arg}(\beta_{GBZ}) = k_m= \frac{2\pi}{N} (m-1), \quad m=1,\cdots,N
\end{equation}
where $N$ is the number of unit cells of the lattice ($N=L/2$).  The non-Bloch
operator $c_{\beta(k_m)}^\dagger$ can be transformed into a new
lattice operator $\tilde{c}_j$ by a Fourier transformation,
\begin{equation}
  \tilde{c}_j^\dagger=\sum_m c_{\beta(k_m)}^\dagger U_{mj}, \quad
  U_{mj}=\frac{1}{\sqrt{N}}e^{-ik_mr_j}.
\end{equation}
After that, we obtain an "artificial'' lattice Hamiltonian,
\begin{equation}
  \widetilde{H}_{PBC} = U^\dagger H(\beta_{GBZ}) U.
  \label{tildeH}
\end{equation}
This type of Hamiltonian was previously studied in Ref. \cite{le.li.20}, which is
named by quasi-reciprocal surrogate Hamiltonian there. The advantage
of Hamiltonian $\widetilde{H}$ is that it is free of non-Hermitian
skin effects, and therefore the concepts, developed for
Hermitian (reciprocal) systems, may also be applicable to it.
In this paper, we are particularly interested in the
behavior of EE on this quasi-reciprocal lattice.

In principle, the quasi-reciprocal Hamiltonian $\widetilde{H}_{PBC}$
gives the OBC spectrum of a system ($H_{OBC}$) in the thermodynamic limit, except
for the isolated topological modes \cite{yo.mu.19}. One can construct
the OBC Hamiltonian $\widetilde{H}_{OBC}$ for the quasi-reciprocal
lattice accordingly. For the bulk energy spectrum, we have
\begin{equation}
\{\epsilon_{OBC}\} \stackrel{L\to\infty}{=}
\{\widetilde{\epsilon}_{OBC}\}
\stackrel{L\to\infty}{=}
\{\widetilde{\epsilon}_{PBC}\}
\end{equation}
When next-nearest-neighbor (NNN) hoppings are included in the
non-Hermitian SSH model, the degree of polynomial $\det[E-H(\beta)]$
is greater than 2, and the GBZ will become non-circular
\cite{ya.wa.18}. In this case, the module $|\beta|$ is
not a constant but depends on the phase angle $k$. By Fourier
transformation, this leads to the long-range hoppings in
quasi-reciprocal lattices \cite{le.li.20}. In practical calculations,
we adopt an approximation to obtain the open boundary of
quasi-reciprocal lattice: we neglect all the hopping terms whose
hopping range is larger than half of the lattice length
($L/2$). This is reasonable since long-range hoppings are
generically power-law decaying \cite{le.li.20}.

A comparison of energy spectra for $H_{OBC}$ and $\widetilde{H}_{OBC}$
is presented in Fig. \ref{fig:gbz}(c) and (d). For circular GBZ, the
quasi-reciprocal lattices possess exactly the same OBC spectra as the
original lattice, no matter what lattice length $L$ is.
This can be understood by examining the structure of $h(\beta)$ in
Eq. (\ref{hbeta}). When $t_3=0$, the module $|\beta|=r$ is a constant,
therefore $h(\beta)$ and $h(k)$ share the same form, except that the
value of $t_2$ is renormalized.
The corresponding quasi-reciprocal
lattices are shown in Fig. \ref{fig:model}(b). In fact,  the
quasi-reciprocal Hamiltonian $\widetilde{H}_{OBC}$ is related to the
original Hamiltonian $H_{OBC}$ by a similarity transformation,
\begin{eqnarray}
  \widetilde{H}_{OBC}=\tilde{S}^{-1}H_{OBC}\tilde{S},
\end{eqnarray}
with $\tilde S=\mathrm{diag}\{r,r,r^2,r^2,\cdots,r^{N-1},r^N,r^N\}$.
This transformation differs from the well-studied similarity
transformation $\bar{H}_{OBC}=\bar{S}^{-1} H_{OBC} \bar{S}$ in Ref. \cite{ya.wa.18} with
$\bar{S}=\mathrm{diag}\{1,r,r,r^2,\cdots,r^{N-1},r^N\}$, which
renormalizes the intracell hopping amplitudes; see
Fig. \ref{fig:model}(c).
Note that $\widetilde{H}_{OBC}$ is also connected to
$\bar{H}_{OBC}$ by a similarity
transformation, $\widetilde{H}_{OBC}=S^{-1}\bar{H}_{OBC} S$, with
$S=\mathrm{diag}\{r,1,r,1,\cdots,r,1\}$. 
Therefore, $\widetilde{H}_{OBC}$ and $\bar{H}_{OBC}$ have the same energy spectra and entanglement entropy.

If $t_3\neq 0$, the long-range hoppings appear in
the quasi-reciprocal lattices due to the non-constant $|\beta|$. There
are tiny difference between the energy spectra of $H_{OBC}$ and
$\widetilde{H}_{OBC}$ [see Fig. \ref{fig:gbz}(d)], for the long-range
hoppings can not be fully taken into account when the length of
lattices is finite. The difference in energy spectra will decrease to
zero in the thermodynamic limit.

\begin{figure}[t]
\includegraphics[width=0.98\linewidth]{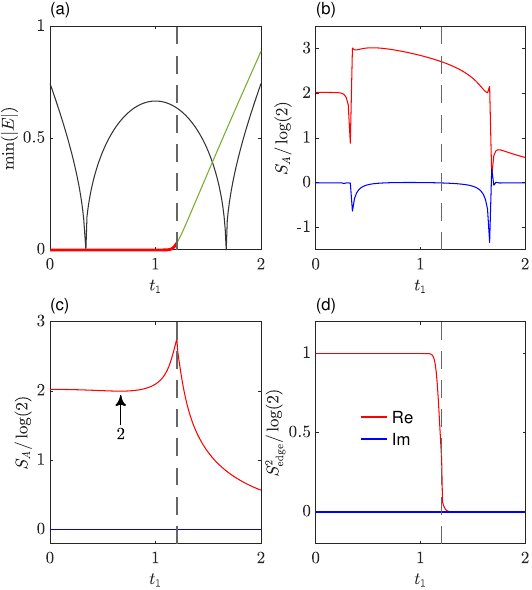}
\caption{\label{fig:t3=0} (a) The minimum of absolute energy spectra
  $|E|$ of the original lattices under PBC (black) and OBC
  (green) conditions. The red line denotes the topological zero-energy
  mode. The von Neumann entropy $S_A$ of the original lattices and the
  quasi-reciprocal lattices are displayed in (b) and (c),
  respectively. Subfigure (d) shows the edge entropy of the
  quasi-reciprocal lattice with $\alpha=2$.  The real part of entropy
  is depicted in red, while the imaginary part is in blue. Parameters
  in the calculations are set to be $t_2=1$, $t_3=0$,  $\gamma=2/3$, and
  $L=100$. The topological transition occurs at
  $t_1\simeq1.202$ and is marked by the black dashed line.}
\end{figure}

\section{Entanglement entropy on Circular GBZ}
\label{section:cirGBZ}

We first present the results of circular GBZ in Fig.~\ref{fig:t3=0},
in which we use the non-Hermitian SSH model with $t_3=0$ as an example. 
It is well-known that the BBC is broken in the non-Hermitian SSH
model: the topological transition of zero-energy modes under OBC can
not be explained by the Bloch Hamiltonian $h(k)$ under PBC
\cite{ya.wa.18,yo.mu.19}; see Fig.~\ref{fig:t3=0}(a).
The entanglement entropy can be calculated by applying the correlation
matrix method described in Sec. \ref{section:bee} to the original
non-Hermitian Hamiltonian under PBC \cite{he.re.19,ch.yo.20}. We show
the results of entanglement entropy for SSH model with $t_3=0$ in
Fig. \ref{fig:t3=0}(b). The entanglement entropy is well-defined in
the line-gap region, \textit{i.e.}, the values are real and
positive-definite (away from the two exceptional points). However, in
the point-gap region, the entanglement entropy is not well-defined: it
becomes complex with a non-zero imaginary part
\cite{he.re.19,ch.pe.22}.  Note that at the special point with
$t_1=t_2$ in the point-gap region, the entanglement entropy is also
real, as studied in Ref. \cite{gu.yu.21}.  In all, the entanglement
entropy defined on the original PBC lattice basically reflects the
characteristics of the Bloch band, and does not respect the BBC.

Hence, we propose to calculate the entanglement entropy using the
quasi-reciprocal lattices defined in Eq. (\ref{tildeH}), which we term
as the entanglement entropy on GBZ or the non-Bloch entanglement entropy.
As displayed in Fig. \ref{fig:t3=0}(c), interestingly, the von Neumann
entropy $S_A$ is well-defined across the parameter space: it is
positive-definite and real with zero imaginary part. It also diverges
at the topological transition points (see the black dashed line), thus
recovering the BBC. The edge entropy, plotted in
Fig. \ref{fig:t3=0}(d), is quantized at 1 (in units of $\log(2)$) within the topological
phase, and changes to zero at the topological transition point.

The entanglement entropy drops to $2\log(2)$ at the point
$t_1 = \gamma$; see Fig.~\ref{fig:t3=0}(c). At this point, the GBZ
collapses into a single point with $r=0$, and the intra-(inter-)unit
cell hoppings of the quasi-reciprocal lattice become $2\gamma$ and 0
(0 and $\infty$), respectively. By performing the similar
transformation $S=\mathrm{diag}\{1,1/r,1,1/r,\cdots,1,1/r\}$, this case can be mapped into the dimerized limit
with intra-unit cell hopping being zero; see Fig. \ref{fig:model}(c). There are
two isolated edge states, hence the entanglement entropy is exactly
$2\log(2)$ \cite{ry.ha.06}.  Note, when $t_1<\gamma$, the Hamiltonian
$H$ cannot be transformed into a Hermitian SSH model
\cite{ya.wa.18}. However, the entropy remains well-defined, as
demonstrated in Fig. \ref{fig:t3=0}(c); see the Appendix \ref{sec:a1}
for the analysis of realness.
This is a significant finding
of the present paper.

\begin{figure}[t]
\includegraphics[width=1.0\linewidth]{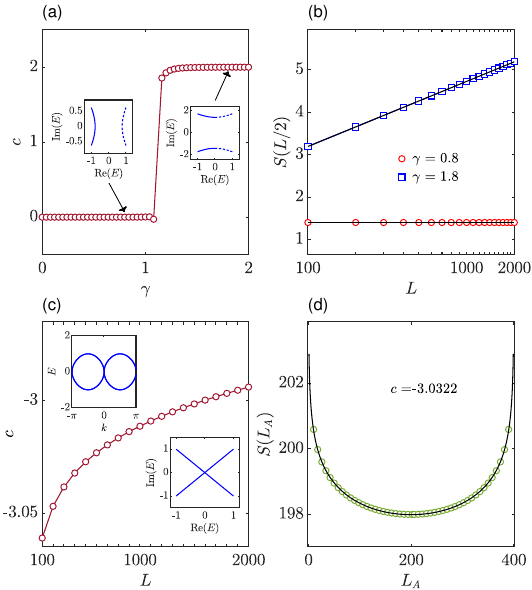}
\caption{\label{fig:epfig} (a) The central charge $c$ of entanglement
  entropy on circular GBZ as a function of
  $\gamma$, excluding the EP at $\gamma=\gamma_c$. The two insets
  display representative energy spectra on two sides of the EP
  $\gamma_c=\sqrt{1.25}$, where the solid (dashed) line indicate the
  occupied (unoccupied) states. (b) The logarithmic scaling of
  entanglement entropy (fixed ratio $L_A/L=1/2$) when $\gamma$ is far
  away from EP.
  (c) The fitted central charge at the EP as a function of total system length
  $L$. In the calculations, $\gamma$ is shifted from
  $\gamma_c$ by a tiny quantity $\eta=1\times 10^{-8}$, and the
  formula in Eq. (\ref{scaling}) is used for the fitting process.
  Insets display the energy spectrum and the dispersion relation,
  revealing two exceptional momenta at $k=0,\pi$ that render the
  momentum-space Hamiltonian defective.  (d) The entanglement entropy
  (green circles) as a function of the subsystem length $L_A$, for
  a fixed length $L = 400$ and at $\gamma=\gamma_c-\eta$.  The black line is the fitting curve
  of entanglement entropy according to Eq. (\ref{scaling}), from which
  the central charge can be extracted.  The hopping parameters for all
  sub-figures are $t_1=0.5$, $t_2=1$, and $t_3=0$.}
\end{figure}

We now investigate the critical behavior of the quasi-reciprocal
lattice Hamiltonian, which exhibits an EP at
$\gamma_c=\sqrt{t_1^2+t_2^2}$. We firstly examine the region away from
the EP ($\gamma\neq\gamma_c$).
In Fig. \ref{fig:epfig}(a), the
fitted central charge $c$ is plotted as a function of $\gamma$ for $t_1=0.5$. For
$\gamma<\gamma_c$, the value of $c$ is zero, indicating that the
system is noncritical in this region. However, for $\gamma>\gamma_c$,
the value of $c$ saturates to 2, implying critical behavior; See
Fig. \ref{fig:epfig}(b) for the representative logarithmic scaling of
entanglement entropy with $L_A/L=1/2$.
The critical behavior is associated with the Fermi surface structure
of the energy spectrum \cite{gu.yu.21}. Since we construct the ground
state according to the real part of energy, the system is an insulator
without any Fermi points when $\gamma<\gamma_c$, but when
$\gamma>\gamma_c$ it becomes a gapless metal with two Fermi points. We
therefore conclude that each Fermi point contributes exactly 1 to the
central charge $c$ of the logarithmic scaling.
This is different from the original SSH model (PBC) studied in
Ref. \cite{gu.yu.21}, where each Fermi point contributes 1/2 to the
value of $c$.

At the EP with $\gamma=\gamma_c$, the entanglement entropy exhibits
divergent behavior. Please refer to the inset in
Fig. \ref{fig:epfig}(c) for the exceptional momentum $k_{EP}$ (0
and $\pi$) in this model, and note that the dispersion of these EPs is
square root dependence on $k$ instead of linear.
To avoid divergence, the entanglement entropy can be calculated by
slightly shifting from the EP $\gamma=\gamma_c-\eta$
\cite{ch.yo.20,tu.tz.22}.
The scaling behavior of entanglement entropy for a fixed system length
$L=400$ is shown in Fig. \ref{fig:epfig}(d), where a tiny 
shift $\eta=1\times 10^{-8}$ is used and $L_A$ is chosen to include
an odd number of unit cells. The extracted central
charge is negative (around $-3$), which is attributed to the exceptional bound state
close to EP \cite{lee.22}. However, the value of central charge
slightly increases when the system length is increasing, as depicted in
Fig. \ref{fig:epfig}(c).

\section{Entanglement entropy on non-circular GBZ}
\label{section:noncirGBZ}

\begin{figure}[t]
\includegraphics[width=1.0\linewidth]{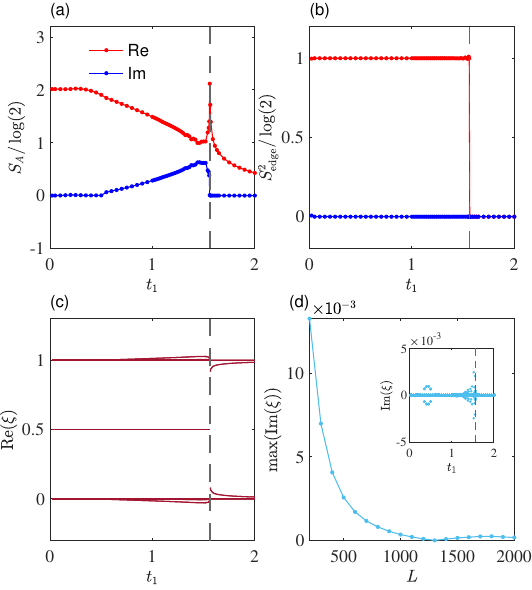}
\caption{\label{fig:t3=02} The von Neumann (a) and $\alpha=2$
  edge (b) entanglement entropy, calculated on the quasi-reciprocal
  lattice, are plotted as a function of $t_1$. Real and imaginary parts of entropy are shown in red and
  blue color, respectively.
  (c) The real part of the entanglement spectrum $\xi$ as a function of
  $t_1$. 
  (d) The maximum imaginary part of entanglement
  spectrum at $t_1=1.45$ varies as a function of system length $L$.
  It  decreases to a small value once $L$ becomes
  sufficiently large.
  The inset shows the imaginary of the entanglement spectrum.
  The topological transition occurs at
  $t_{1c}=1.562$ and is denoted by the black dashed line.
  The values of parameters used in the calculations are $\gamma=2/3$,
  $t_2=1$, $t_3=0.2$, and $L=800$.}
\end{figure}

\begin{figure}[t]
\includegraphics[width=0.90\linewidth]{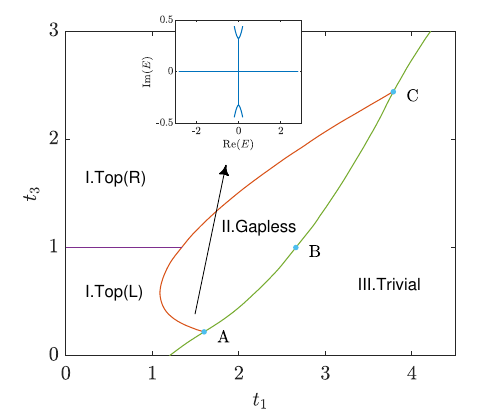}
\caption{\label{fig:phase} Phase diagram of the non-Hermitian SSH
  model under OBC in the $t_1-t_3$ plane. Phase I is the gapped
  topological phase.
  The letters $L$ and $R$, abbreviating "left" and "right", indicate
  the accumulating direction of skin modes.
Phase II is the gapless semimetal phase, with the gap closing at some
exceptional momentum points. A typical energy spectrum is shown in the
inset. The phase boundaries around Phase II are
all exceptional. Phase III is the gapped topologically
trivial phase.
  The
  parameters used in the calculations are $\gamma=2/3$, $t_2=1$, $L=200$.}
\end{figure}

We now consider the general cases of non-circular GBZ, where
long-range hopping appears in the quasi-reciprocal lattices. The
entanglement entropy calculated on the quasi-reciprocal lattices is
presented in Fig. \ref{fig:t3=02} for the case of $t_3=0.2$.
The von Neumann entropy exhibits singularity at the topological
transition point, similar to the circular GBZ case in Fig.
\ref{fig:t3=0}(c). However, it is not well-defined (\textit{i.e.},
becomes complex valued) anymore in the topological
region with $t_1<t_{1c}$.
As shown in Fig. \ref{fig:t3=02}(c), the values of entanglement
spectrum go outside the bound $[0,1]$, which leads to an imaginary part
in the von Neumann entanglement entropy.

However, the behavior of the edge entanglement entropy is significant: as
shown in Fig. \ref{fig:t3=02}(b), it retains real ($L\to \infty$), quantized values of
1 in the topological region, but then drops to zero upon entering the
topologically trivial region.
This means that the edge entanglement entropy is a good topological
indicator, able to recover the BBC, even in the case of non-circular
GBZ.

The phase diagram of quasi-reciprocal Hamiltonian is shown in
Fig. \ref{fig:phase}. A gapless semimetal phase (Phase II) appears
when $t_3\neq 0$ \cite{yo.mu.20}, in which the system becomes exceptional.  We then
consider possible scaling behaviors along the boundaries delineating
this exceptional phase \cite{yi.ha.23}.
Since the von Neumann entropy $S_A$ is complex, we adopt the
generalized entanglement entropy $S_A^g$, proposed in
Ref. \cite{tu.tz.22}, to investigate the critical properties,
\begin{equation}
  S_A^g=-\mathrm{tr}_B(\rho^{\text{RL}}_A\log(|\rho^{\text{RL}}_A|)),
\end{equation}
which in terms of the entanglement spectrum can be written as
\begin{equation}
  S_A^g=-\sum_l(\xi_l\log(|\xi_l|)+(1-\xi_l)\log(|1-\xi_l|)).
\end{equation}
The central charges can be obtained as usual by fitting the
generalized entropy $S_A^g$ to the scaling formulas given
in Eqs. (\ref{scaling}) and (\ref{scaling2}).

\begin{figure}[t]
\includegraphics[width=1.0\linewidth]{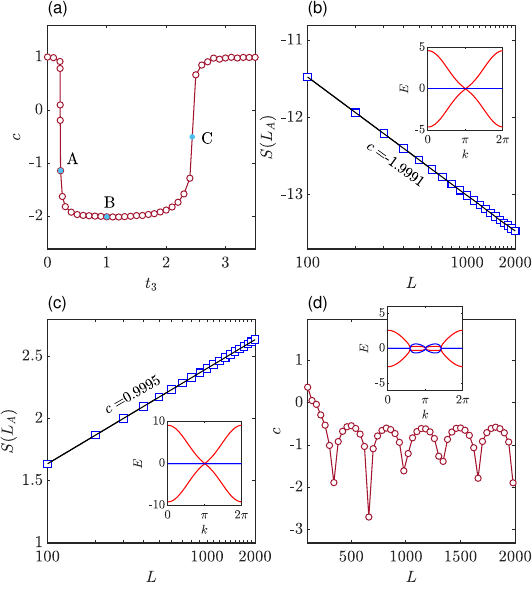}
\caption{\label{fig:rbep} (a)
  The fitted central charges are plotted along the green line in
  Fig. \ref{fig:phase}, which delineates the boundaries between the
  topological phase (the semimetal phase) and the trivial phase.
  The parameters used in the calculations are $\gamma=2/3$, $t_2=1$, $L=200$.
  (b) The logarithmic scaling of generalized
  entanglement entropy at point $B$ in Fig. \ref{fig:phase} with
  $t_1=8/3$ and $t_3=t_2=1$.
  (c) The logarithmic scaling of generalized
  entanglement entropy at $t_1=4.65725241$
  and $t_3=3.5$. (d) The fitted cental charge as a function of system length $L$
  at $t_1=1.08907404$ and $t_3=0.6$ (on the topological-semimetal phase
  boundary).
  In all sub-figures, a momentum shift $\delta=1\times 10^{-4}$ is
  used in the calculations. 
  The insets in (b), (c), and (d) show the dispersion relations, where
  the real and imaginary part of energy are
  plotted in red and blue color, respectively.  
}
\end{figure}

To avoid the singularity at $k_{EP}$, we introduce a tiny momentum shift
$\delta$ \cite{tu.tz.22} and choose the momentum as $k=k_m+\delta/N$,
with $k_m$ being the ordinary momentum in Eq. (\ref{km}).
The fitted central charges along the topological-trivial 
and the semimetal-trivial phase boundaries (the green line in Fig. \ref{fig:phase}) are plotted in
Fig. \ref{fig:rbep}(a), where a tiny shift $\delta=1\times 10^{-4}$ is used.
For $t_3=0$, the quasi-reciprocal model can be mapped to the Hermitian
SSH model (see Fig. \ref{fig:model}(c)), so that the central charge at
the topological transition point is $c=1$.  The value of $c$ decreases
slightly from 1 when approaching the tri-critical point $A$.
Upon entering the exceptional region, the value of $c$ drops to a
negative value (around $-2$), which is governed by the exceptional bound
state \cite{lee.22}.
At the critical point $B$ with $t_3=t_2=1$ and $t_1=2+\gamma$,
the GBZ is a circle with $|\beta|=1$. The momentum-space Hamiltonian in the vicinity of the
exceptional momentum $k_{EP}=\pi$ can be written as
\begin{equation}
  H(k_{EP}+k)=\left(\begin{matrix}0&2\gamma+k^2\\
                      k^2& 0 \end{matrix}\right),
\end{equation}
which exhibits a linear dispersion. This Hamiltonian is a minimal exceptional bound
state model \cite{lee.22}, and the central charge is precisely $-2$ at
this point \cite{tu.tz.22,lee.22}, as illustrated in
Fig. \ref{fig:rbep}(b) with $\delta=1\times 10^{-4}$. For the other points on the semimetal-trivial phase boundary
different from point $B$, the dispersion deviates from being exactly
linear, therefore the central charge deviates from the value of $-2$
(\textit{i.e.}, not an integer); see Fig. \ref{fig:rbep}(a).
When passing through the tri-critical point $C$ along the phase boundary,
the central charge becomes positive ($c\approx 1$) again; see
Fig. \ref{fig:rbep}(c).

The situation of the topological-semimetal phase boundary (red line in
Fig. \ref{fig:phase}) is more intricate. On the boundary delimiting
the topologically trivial phase (the green line in
Fig. \ref{fig:phase}), the exceptional momentum $k_{EP}$ is 
at $\pi$. However, on the topological semimetal phase boundary, two
additional exceptional momenta different from $0$ or $\pi$ appear, as
depicted in the inset of Fig. \ref{fig:rbep}(d).
Depending on the length of the system, it is possible that the crystal
momentum does not coincide with $k_{EP}$. This renders the calculation
of the central charge challenging.  For a given length $L$, we employ
the formula of Eq. (\ref{scaling}) to obtain the fitted central charges,
and then plot them as a function of $L$ in
Fig. \ref{fig:rbep}(d). Evidently, the fitted central charge $c$ exhibits an
oscillatory behavior.

\section{summary and discussion}
\label{section:summary}

We introduce the quasi-reciprocal Hamiltonian and lattice in real
space, which is obtained by performing a regular Fourier
transformation on the non-Bloch Hamiltonian. We calculate the
entanglement entropy on these quasi-reciprocal lattices, and term it
as the non-Bloch entanglement entropy.  Since the quasi-reciprocal
lattice is free of skin effect, the non-Bloch entanglement entropy
exhibits well-behaved characteristics.  For cases where the GBZ is
circular, the non-Bloch entanglement entropy is shown to be
well-defined (real and positive-definite).  The non-Bloch entanglement
entropy exhibits a logarithmic scaling in the critical region. It is
found that the critical behavior is associated with the Fermi surface
structure of the energy spectrum: each Fermi point contributes
precisely 1 to the central charge $c$ of the logarithmic scaling. At
the exceptional point, the central charge becomes negative due to the
presence of the exceptional bound state.  When the GBZ is
non-circular, the non-Bloch von Neumann entropy becomes complex-valued
in the topological region, due to the emergence of long-range hopping
in the quasi-reciprocal lattice. Nonetheless, the non-Bloch edge
entanglement entropy remains real and quantized, thus serving as a
reliable topological indicator recovering the BBC.
Furthermore, We compute the topological phase diagram and reveal the
critical behavior along exceptional phase boundaries between
topological, semimetal, and trivial phases. The central charge
exhibits interesting behaviors, such as becoming precisely $-2$ at the
critical point with $t_3=t_2$ or oscillating as a function of the
system's length on the topological-semimetal phase boundary.
Overall, this study sheds new light on the physical meanings of the
GBZ and non-Hermitian entanglement entropy, opening new possibilities for
studying many-body physics based on the GBZ. 

Finally, we emphasize that the non-Bloch entanglement entropy
accurately characterizes the entanglement behavior of the original
non-Hermitian systems under OBC.
Motivated by the successful measurement of entanglement entropy in
Hermitian phononic systems \cite{li.zh.24}, we anticipate that the
non-Bloch entanglement entropy can be experimentally identified and
measured in future endeavors.

\section*{ACKNOWLEDGMENT}
We are grateful for the discussions with Zhesen Yang, Hantao Lu, and
Shijie Hu. This work is supported
by the National Natural Science Foundation of China (Grants
No. 11974293).

\textit{Note added}.
After completing our work, we became aware of a recent
paper \cite{mu.ar.24}. Some of their findings overlap with our results.

\appendix

\section{The realness of entanglement entropy for circular GBZ}\label{sec:a1}

The correlation matrix $C$ is a projection that projects onto the
occupied band: $C^2=C$. However, the truncated correlation matrix
$C^A$, which is related to the entanglement spectrum, is not a
projection \cite{le.ye.15,lee.22}. We therefore focus on the quantity
$C_{\mathrm{mix}}=C_A-C_A^2$ to investigate the realness of entanglement
entropy \cite{lee.22}. An eigenstate $|\psi\rangle$
of $C_{A}$ with eigenvalue $\xi$ is also an eigenstate of
$C_{\mathrm{mix}}$ with eigenvalue $\xi(1-\xi)$. For a single entanglement
cut, the correlation matrix $C$ can be decomposed into four parts
\begin{equation}
    C=\left(\begin{matrix}
        C^A&C^{AB}\\C^{BA}&C^B
    \end{matrix}\right).
\end{equation}
Introducing a real-space projector $P_A$ onto subsystem $A$
\begin{equation}
    P_A=\left(\begin{matrix}
        1_{A}&0\\0&0
    \end{matrix}\right)
\end{equation}
it is straightforward to show that
\begin{equation}
    (P_ACP_A)^2 - P_ACP_A = P_AC[P_A,C]P_A,
\end{equation}
that is,
\begin{align}
  \left(\begin{matrix}
        C_A^2-C_A&0\\0&0
    \end{matrix}\right)
 =\left(\begin{matrix}
        -C_{AB}C_{BA}&0\\0&0
    \end{matrix}\right)
\end{align}
Therefore, we obtain $C_{\mathrm{mix}}=C_{AB}C_{BA}$.

We consider the two-band non-Bloch Hamiltonian in Eq. (\ref{hbeta})
with circular GBZ (\textit{i.e.}, $t_3=0$),
\begin{align}
    h(\beta)&=\left(\begin{matrix}0&h_+(\beta_k)\\h_-(\beta_k)&0\end{matrix}\right)
\end{align}
where $h_+(\beta_k)=(t_1+\gamma)+t_2e^{-\mathrm{i}k}/r$,
$h_-(\beta_k)=(t_1-\gamma)+t_2r e^{\mathrm{i}k}$, and
$|\beta_k|=r=\sqrt{|(t_1-\gamma)/(t_1+\gamma)|}$ is a constant.  For a
half-filled system, correlation matrix in momentum-space
representation is given by \cite{lee.22}
\begin{align}
  P(k)&=\frac{1}{2}\left(
        \begin{matrix}1&-h_+(\beta_k)/\varepsilon(\beta_k)\\-h_-(\beta_k)/\varepsilon(\beta_k)&1 \end{matrix}\right),
\end{align}
where $\varepsilon(\beta_k)=\sqrt{h_+(\beta_k)h_-(\beta_k)}$ is
the positive root.
The block element of correlation matrix in real space can be obtained by a
Fourier transformation \cite{lee.22},
\begin{align}
C_{ij}&=\langle x_i|P|x_j\rangle =\frac{1}{N}\sum_k e^{\mathrm{i}k(x_i-x_j)} P(k)\nonumber\\
          &= \left(\begin{matrix}  \frac{1}{2} \delta_{x_i,x_j} & f_+(x_i-x_j) \\
                    f_-(x_i-x_j) &  \frac{1}{2} \delta_{x_i,x_j} \end{matrix}\right),
\end{align}
where
\begin{align}\label{for:appe1}
f_{\pm}(x)=-\frac{1}{2N}\sum_k e^{\mathrm{i}kx}\frac{h_{\pm}(\beta_k)}{\varepsilon(\beta_k)},
\end{align}
and $N$ is the number of unit cells. Note that $f_{\pm}(x)$ is
always a real
number for any value of $x$, since
\begin{align}\label{for:appe2}
  \frac{h_\pm(\beta_{k})}{\varepsilon(\beta_k)} \equiv
  \Big[\frac{h_\pm(\beta_{-k})}{\varepsilon(\beta_{-k})} \Big]^*
\end{align}
Then, the matrix
$C_{\mathrm{mix}}$ can be expressed as
\begin{align}\label{for:cmix}
(C_{\mathrm{mix}})_{x_1,x_2} & = \sum_{x_3\in B} C_{x_1,x_3}C_{x_3,x_2} 
          = \left(\begin{matrix}
                    U & 0 \\
                    0 & V 
                  \end{matrix}\right) \nonumber\\
            U&= \sum_{x_3\in B} f_+(x_1-x_3)f_-(x_3-x_2)\nonumber\\
            V&= \sum_{x_3\in B} f_-(x_1-x_3)f_+(x_3-x_2),
\end{align}
in which $x_1$ and $x_2$ denote the sites in subsystem $A$, while
$x_3$ denotes the sites in subsystem $B$. Clearly, $C_{\mathrm{mix}}$ is a real matrix.

\subsection{Case I: $t_1>\gamma$}
In this section, we demonstrate that $C_{\mathrm{mix}}$ is a Hermitian matrix
with real eigenvalues in the region of $t_1>\gamma$. For $t_1>\gamma$,
the following two conditions are satisfied,
\begin{align}\label{for:appe3}
{h_+(\beta_{k_1})}{h_-(\beta_{k_2})}\equiv {h_+(\beta_{-k_2})}{h_-(\beta_{-k_1})}
\end{align}
\begin{align}
\varepsilon(\beta_{k_1})\varepsilon(\beta_{k_2})\equiv\varepsilon(\beta_{-k_2})\varepsilon(\beta_{-k_1}).
\end{align}
Therefore, we have
\begin{align}\label{for:appe4}
  \frac{h_+(\beta_{k_1})}{\varepsilon(\beta_{k_1})}\frac{h_-(\beta_{k_2})}{\varepsilon(\beta_{k_2})}
  \equiv
  \frac{h_+(\beta_{-k_2})}{\varepsilon(\beta_{-k_2})}\frac{h_-(\beta_{-k_1})}{\varepsilon(\beta_{-k_1})}
\end{align}
Combining Eq. (\ref{for:appe4}) with the definition of $f_{\pm}(x)$ in
Eq. (\ref{for:appe1}), we obtain,
\begin{align}\label{for:appe5}
f_+(y_1)f_-(y_2)\equiv f_+(-y_2)f_-(-y_1),
\end{align} 
where $y_1$ and $y_2$ denote any positions on the
lattice. By the relation in Eq. (\ref{for:appe5}), it is straightforward to
show that
\begin{align}
(C_{\mathrm{mix}})_{x_1,x_2}=(C_{\mathrm{mix}})_{x_2,x_1}.
\end{align}
Considering that $C_{\mathrm{mix}}$ is real, we thus conclude that
$C_\mathrm{mix}$ is Hermitian when $t_1>\gamma$. This validates the
realness of the entanglement entropy in this region.

\subsection{Case II: $t_1< \gamma$}

In the region $t_1<\gamma$, the non-Hermitian SSH model without $t_3$
can not be transformed into a Hermitian one through a similarity
transformation \cite{ya.wa.18}. In this case, the
Eq. (\ref{for:appe3}) is no longer valid, and $C_{\mathrm{mix}}$ is no
longer Hermitian.  In the following, we illustrate the general
properties of $C_{\mathrm{mix}}$ matrix using a simple example where
the lattice has only four unit cells ($N=4$).  When $t_1<\gamma$, the
general structure of $C_{\mathrm{mix}}$ is the sum of a
skew-symmetric matrix and a diagonal matrix. For the $N=4$ case, it is
\begin{align}
  C_{\mathrm{mix}}=
    \left(\begin{matrix}a&0&c&0\\0&b&0&c\\- c&0&b&0\\0&- c&0&a\end{matrix}\right),
\end{align}
which contains three different variables
\begin{align}
    a&=f_+(2)f_-(2)+f_+(1)f_-(3)\\
    b&=f_+(2)f_-(2)+f_+(3)f_-(1)\\
    c&=f_+(3)f_-(2)+f_+(2)f_-(3)
\end{align}
The characteristic equation of $C_{\mathrm{mix}}$ is
\begin{align}
  \det(\lambda-C_{\mathrm{mix}})=
  \Big[\lambda^2-(a+b)\lambda+(ab+c^2)\Big]^2=0
\end{align}
To solve this, we 
observe that the following auxiliary function
\begin{align}
g(k_1,k_2)=\frac{h_+(k_1)h_-(k_2)}{\varepsilon(k_1)\varepsilon(k_2)}
\end{align}
takes values as in the table below
\begin{center}
\noindent
\begin{tabular}{ |c|c|c|c|c|  }
 \hline
 \backslashbox{$k_1$}{$k_2$}&$0$&$\pi/2$&$\pi$&3$\pi/2$\\
  \hline
 $0$&$1$&$z$&$z^2$&$z^*$\\
  \hline
 $\pi/2$&$1/z$&$1$&$z$&$-1$\\
  \hline
 $\pi$&$1/z^2$&$1/z$&1&$1/z^*$\\
  \hline
 $3\pi/2$&$1/z^*$&$-1$&$z^*$&$1$\\
 \hline
\end{tabular}
\end{center}
This table contains only one independent variable, namely
$z=g(0,\pi/2)$. Combining this table and the definition in
Eq. (\ref{for:appe1}), we obtain
\begin{align}
    a+b&=\frac{1}{4}\\
    ab+c^2&=0
\end{align}
which holds except at the exceptional point. Therefore, the eigenvalues
$\lambda$ of $C_{\mathrm{mix}}$ are $0,0,1/4,1/4$, resulting in a real
entanglement entropy with a precise value of $2\log(2)$.  When $N>4$,
the above table becomes much more complex. However, we numerically
verify that the entanglement entropy is always real.


%

\end{document}